\documentclass[twocolumn]{sig-alt-hotnets}
\usepackage{subfigure} 
\usepackage{graphicx}
\usepackage[hyphens]{url}
\usepackage{color}
\usepackage{sgame}

\begin{document}

\title{Architectural innovation: A game-theoretic approach}

\author{Ioannis Avramopoulos\thanks{The author is with the National Technical University of Athens. His email is {\texttt ioannis.avramopoulos@gmail.com}.}}

\maketitle

\thispagestyle{empty} 

\newtheorem{definition}{Definition}
\newtheorem{proposition}{Proposition}
\newtheorem{theorem}{Theorem}
\newtheorem{corollary}{Corollary}
\newtheorem{lemma}{Lemma}

\begin{abstract}
In this paper, we view the Internet under a game-theoretic lens in an effort to explain and overcome the Internet's innovation slump. Game Theory is used to model Internet environments as problems of {\em technological competition} toward the end of understanding their emergent phenomena and the evolutionary forces that shape them. However, our results extend beyond understanding the Internet architecture toward helping the Internet population achieve socially desirable outcomes.
\end{abstract}

\section{Introduction}

Scott Shenker has been quoted to say that ``The Internet is an equilibrium, we just have to identify the game''~\cite{Algorithmic-Game-Theory}. Our contribution in this paper is to that end. We start with the empirical evidence that the Internet is in fact an equilibrium, which is no other than the {\em Internet impasse}~\cite{Impasse}. `Is your system incrementally deployable?' has perhaps become the most frequently asked question in networking for a good reason: Innovation in the Internet has stagnated~\cite{Naughton}, and many attribute this stagnation to the lack of incremental deployability of innovative technologies, a problem that our network architecture exacerbates. This has spurred research efforts toward devising incrementally deployable systems  as well as architectures enabling favorable to incremental deployability environments. The overall research effort is colossal, and perhaps Shenker prompted the community to identify the Internet's game for a good reason, namely, to end the Internet's bane.

In this paper, we embark on a scientific exploration of network architecture and architectural evolution whose ultimate destination is a radical new method for the deployment of superior technologies at the network and transport layers of the Internet. The method is that of {\em insuring deployment investments}, however, to explain why this method is promising requires a non-trivial mathematical apparatus and significant preparation. Along the way, we lay a mathematical foundation for network architecture by formalizing the process of architectural evolution at the network and transport layers of the Internet (the narrow ``waist'' of the Internet hourglass). What do we mean by the terms {\em network architecture} and {\em architectural evolution} though?

\subsection{Network architecture}

Whenever researchers and practitioners alike talk about the Internet architecture they refer to a system of technologies that have received adoption as well as their overarching technical principles. We believe that this technology-centric perspective has been harmful to the networking community's effort to effect innovation. This effort has been focused on facilitating the {\em implementation} of innovation in software and hardware through designing an Internet around the concept of open interfaces. Although implementing innovation is clearly an important step in effecting it, it barely touches upon the thornier {\em human-centric} aspects of this process: The {\em availability} of superior technologies is not a sufficient condition for their adoption, far from it.

In this paper, we propose a change in perspective from the technology-centric view toward a human-centric one. In the new perspective, the Internet is understood as a {\em society of agents} who choose among competing technologies and refer to network architecture as {\em the collective outcome of choices that the agents in the society make as well as the principles that govern the society's decisions}. That is not to say that we should abandon the former perspective, in fact, they are complementary, and our definition reflects the judgment call that the human-centric aspects of network architecture are poorly understood.

\subsection{Architectural evolution}

Similarly, there are two conceivable (by us) ways to define architectural evolution, an {\em endoscopic} and a {\em macroscopic} one. In its endoscopic definition, architectural evolution is the process by which new Internet technologies emerge from previous ones. In its macroscopic definition, it is the process by which new  technologies are introduced and spread in the Internet's ecosystem (that is, how they {\em diffuse}). We believe that the niche of the endoscopic approach is to understand {\em how technologies are created} whereas the niche of the macroscopic one is to understand {\em how innovation is effected}. In the words of W. Brian Arthur, ``The people who have thought hardest about the general questions of technology have mostly been social scientists and philosophers, and understandably they have tended to view technology from the outside as stand-alone objects. \ldots \hspace{0.2 mm} Seeing technology this way, from the outside, works well enough if we want to know how technologies enter the economy and spread within it''~\cite{Technology}.

We believe that the bottleneck in Internet innovation is not a lack of creativity in devising new technologies but rather a lack of creativity in profitably deploying them, and, therefore, in this paper, we take on the macroscopic approach to understanding architectural evolution. To the extent of our knowledge this is the first attempt to understand the principles of the diffusion of architectural innovation, although the diffusion of innovation in society, in general, has been the subject of extensive study~\cite{Diffusion}. Innovating at the core of the Internet architecture is unique in that the outcome is shaped by the presence of strong externalities.\footnote{This paper is about innovation whose success depends on {\em positive externalities}. The study of innovation whose success depends on {\em managing negative externalities} is the subject of an in-preparation paper by the same author.}

\subsection{Game Theory}

Viewed as a social system, the Internet is, to a first approximation, an {\em anarchy}, that is, a society without a publicly enforced government or violently enforced political authority,\footnote{\url{http://en.wikipedia.org/wiki/Anarchy}} in the sense that its agents are free to choose the technologies they employ. Because of the absence of a central authority, the social outcome is determined by strategic interaction, and, therefore, the approach to studying such a society is naturally {\em game-theoretic} as Game Theory's raison d'\^etre is to study human behavior in strategic environments.

Going further, we propose a simple representation of this social system as a mathematical game in which {\em the society's agents correspond to players}, {\em technologies correspond to strategies}, and {\em payoffs capture the incentive structure of the society}. The outcome of such a game (typically a {\em Nash equilibrium}) becomes the social outcome, which, according to our earlier definition, is the network architecture itself. In prompting the community to identify the Internet's game, was Shenker really prompting the community to identify the Internet's architecture?

\subsubsection*{Effecting innovation}

The Internet-as-an-equilibrium paradigm not only extends the theoretical understanding of network architecture, but it has the practical benefit of lending itself to an approach for {\em effecting innovation:} To effect innovation all we need to do is {\em change the Internet's equilibrium} in favor of innovative technologies. In this way, meeting Shenker's challenge becomes essential to moving the Internet forward, and, in fact, we show that insurance works in exactly this way.

\subsection{Overview of our results}

\subsubsection{Scope of this paper}

The Internet has enabled significant innovation in society at large, however, it has resisted repeated efforts to effect innovation at its core architecture, namely, the protocols running at the network and transport layers of the Internet (i.e., layers 3 and 4 in the OSI reference model). This core architecture has facilitated significant innovation at the link layer (e.g., the rapid transition to 3G and 4G systems in wireless and the rapid transition to optical technologies for fixed access) as well as the application layer (e.g., Google, Facebook, BitTorrent), however, it is increasingly being held responsible for stifling the emergence of radical new applications~\cite{Naughton}.

In this paper, our goal is twofold, namely, to both {\em explain} why innovation is failing and to {\em propose a course of action} on how innovation can be effected. Our inquiry on architectural evolution spans both the network and the transport layers, as to a large extent they are inseparable pieces of the core architecture. As part of our inquiry, we expend much effort on understanding the {\em evolutionary forces} that shape the core architecture. For example, a question of great concern is whether players are {\em myopic} in which case incremental deployability would {\em characterize} deployable innovation.

\subsubsection{Understanding architectural evolution}

Our methodology to understanding architectural evolution consists of devising mathematical models and using game-theoretic reasoning to explain what we believe are the most pertinent to architectural innovation phenomena, namely, TCP's dominance at the transport layer over more aggressive transport protocols and the failure to effect innovation at the network and transport layers. That TCP dominates aggressive transport protocols is hardly an ``innovation failure,'' however, the explanation sheds important light on the evolutionary forces at work (ruling out a myopic model of the Internet population) and facilitates the analysis on why TCP has not yet been dominated by less aggressive but more efficient variants (such as TCP Vegas~\cite{TCP-Vegas}). 

Toward the end of understanding why TCP dominates the transport layer, in Section~\ref{pwoeriudffmgn}, we study a mathematical abstraction modeling technological competition at this layer. This abstraction is an important step toward answering Papadimitriou's conjecture that TCP/IP congestion control is a Nash equilibrium~\cite{Internet-Game-Theory}, and it furthers the understanding of the Internet in two fundamental ways: First, it demonstrates the actuality of {\em cooperation} as an evolutionary force in the Internet. Second, it informs the networking community's effort to effect innovation by shedding light on the concept of incremental deployability: That TCP is an equilibrium implies that it has been able to outcompete its more aggressive (and, therefore, incrementally deployable) relatives. This, in turn, implies that incremental deployability is neither sufficient nor necessary for a technology to receive deployment (and, therefore, that it is not panacea in the efforts to effect innovation contrary to folk wisdom).

Thereafter, in Section~\ref{iwueyrajshdgflgkhj}, we study a second (but related) abstraction whose aim is to capture and explain a recurring phenomenon in the Internet according to which superior over the status quo transport-layer and network-layer architectures emerge whose adoption fails. These technologies are notoriously difficult to deploy incrementally as their success is contingent on adoption by a significant number of peers. However, the reasons that such dependency causes adoption to fail are not well understood. Our finding is that a plausible explanation for these recurring adoption failures is  {\em risk aversion}, that is, a propensity in human nature to avoid risky undertakings even if in expectation they are beneficial: Since success is contingent on significant adoption, unilateral investment in these technologies entails the risk of being futile (unless other players join forces), and although adoption is the rational outcome, {\em uncertainty} dominates the players' decisions leading them to desist from the investment effort.

\subsubsection{Engineering architectural evolution}

Finally, in Section~\ref{qwioerqwerjhejkl}, we take on the problem of fostering achievement of socially desirable outcomes. To that end, we propose a method by which a superior technology whose success depends on positive externalities can overcome a cooperation breakdown. The idea in this deployment method is to conciliate aversion against risk by offering the possibility of {\em insurance}, which, as it turns out, in principle, induces adoption {\em without} the purchase of insurance by any potential adopter. The method's thrust is to change the incentive structure of the environment: Purchasing insurance becomes an incrementally deployable strategy against staying with the incumbent and adopting the new architecture without purchasing insurance becomes incrementally deployable against the purchase of insurance, the end outcome being that the incumbent (and possibly the purchase of insurance) is leapfrogged by the new architecture.






\section{Game Theory background}

This section reviews standard material in Game Theory~\cite{Myerson, Fudenberg, Taylor}.

\subsection{Games in strategic form}

\subsubsection{Players, strategies, and payoffs}

We begin with the definition of games in {\em strategic form}. To define a game in this form, we need to specify the set of players, the set of strategies available to each player, and a utility function for each player defined over all possible combinations of strategies that determines a player's payoffs. Formally, a  strategic-form game $\Gamma$ is a triple $$\Gamma = (I, (S_i)_{i \in I}, (u_i)_{i \in I}),$$ where $I$ is the set of players, $S_i$ is the set of pure strategies available to player $i$, and $u_i: S \rightarrow \mathbb{R}$ is the utility function of player $i$ where $S = \times_i S_i$ is the set of all strategy profiles (combinations of strategies).

\subsubsection{Notational convention}

Let $N$ be the number of players. We often wish to vary the strategy of a single player while holding other players' strategies fixed. To that end, we let $s_{-i} \in S_{-i}$ denote a strategy selection for all players but $i$, and write $(s'_i, s_{-i})$ for the profile
\begin{align*}
(s_1,\ldots,s_{i-1}, s'_i,s_{i+1},\ldots, s_N).
\end{align*}

\subsubsection{Nash equilibrium}

A pure strategy profile $\sigma^*$ is a {\em Nash equilibrium} if, for all players $i$,
\begin{align*}
u_i(\sigma_i^*, \sigma_{-i}^*) \geq u_i(s_i, \sigma_{-i}^*) \text{ for all } s_i \in S_i.
\end{align*}
That is, a Nash equilibrium is a strategy profile such that no player can obtain a larger payoff using a different strategy while the other players' strategies remain the same.

\subsection{Supergames}

The analysis of strategic-form games depends on the assumption that players choose their strategies {\em simultaneously} and that their choices cannot change post hoc. {\em Repeated games} (also called {\em supergames}) allow instead for the possibility of strategy revisions.

We consider supergames that are sequences of identical {\em stage games}, i.e., simultaneous-move games in strategic form. Let $$\Gamma = (I, (A_i)_{i \in I}, (v_i)_{i \in I})$$ be the stage game, where $A_i$ is {\em action space} of player $i$ and $v_i$ is her stage-game utility function. To define the supergame, we must specify the players' set of strategies and their utility functions.

\subsubsection{Strategies}

We consider supergames in which players observe their opponents' actions at the end of each period, and condition their choices on their opponents' history of play up to the current period. Let $a^t = (a^t_1, \ldots, a^t_N)$ be the actions that are played in period $t$, let $h^t = (a^0, a^1, \ldots, a^{t-1})$ be the history of play up to period $t$, and let $H^t$ be the space of all period-$t$ histories. A pure strategy $s_i$ for player $i$ is a sequence $\{ s_i^t \}$ whose elements $s_i^t$ are maps from period-$t$ histories $h^t \in H^t$ to actions $a_i \in A_i$.

\subsubsection{Utility functions}

We make the standard assumption that payoffs are discounted exponentially and that the discount factors $\alpha_i$ (where $0< \alpha_i < 1$) remain constant through time. Then given a strategy profile $$s = (\{ s_1^t \}, \ldots, \{ s_N^t \}),$$ the payoff to player $i$ is
\begin{align*}
u_i(s) = \sum_{t=0}^{\infty} \alpha^t_i v_i(s^t(h^t))
\end{align*}
where $$s^t = (s_1^t, \ldots s_N^t).$$
Since $0 < \alpha_i < 1$ this infinite series converges for any sequence of payoffs provided these payoffs are finite.

\subsubsection{Nash equilibrium}

Finally, the previous definition of the Nash equilibrium continues to hold in the setting of supergames.

\section{Why is TCP dominant?}
\label{pwoeriudffmgn}

TCP congestion control was designed and deployed in the Internet in response to a congestion collapse in 1986~\cite{Jacobson} and has remained in stable operation since then. The attempts to effect innovation at the transport layer fall into two classes, those that promote TCP-friendly technologies (e.g.~\cite{TCP-Vegas, FAST-TCP}) and those that promote radical digressions (e.g., Skype). In this section, we are concerned with the latter type of innovation, and, in particular, the effortless strategy of obviating congestion control that, on the surface, appears to be the most profitable one as well. Understanding why TCP has resisted invasion by technologies of this sort has been posed by Papadimitriou as one of the most significant open problems in Internet-related algorithmic research: ``This ingeniously simple scheme seems to work, and its users do not seem eager to abandon it for something more aggressive, but the origins of this apparent success and acquiescence are not well understood. One is justified to wonder: {\em Of which game is TCP/IP congestion control the Nash equilibrium?}''~\cite{Internet-Game-Theory}. 

In this section, we attempt to answer this question by identifying a game where {\em conditional} use of TCP congestion control by all players is an equilibrium. Cooperation is conditional in that players are eager to obviate congestion control to penalize defectors, which implies that defection does not pay off in the long run. Since {\em unconditional cooperation} can be exploited by defectors and is, therefore, not an equilibrium, we may conclude that TCP congestion control's architectural stability is imputable to player's having a {\em theory of mind}. 

In this section, we draw on Taylor's analysis of {\em symmetric prisoner's dilemma supergames}~\cite{Taylor}; our contribution is limited to showing that these games are an apt model of Internet congestion control and to interpreting the ensuing implications in the context of Internet architecture.

\subsection{TCP Background}

To transfer data in the Internet pairs of endhosts establish {\em transport sessions}. Each transport session has a source, a destination, and a path in the network, which is a sequence of links. Each link has a capacity, which must be shared among the competing sessions crossing that link. TCP congestion control is a distributed algorithm for allocating {\em bitrates} to transport sessions. These rates are a solution to the {\em rate allocation problem}, which is to allocate bitrates for all transport sessions that are feasible (in that the capacity constraints are satisfied) so that the allocation is both {\em efficient} (in that network capacity is not wasted) and {\em fair} (in that sessions are treated ``equally''). 

The allocations that have received most attention in the literature are {\em max-min fair} allocations~\cite{Bertsekas-Gallager} and {\em proportionally fair} ones~\cite{Kelly1}. In a max-min fair allocation, all sessions are entitled to the same share of their bottleneck link. However, if a session cannot use all of its share, perhaps because it has a slower rate in another bottleneck, then the excess capacity is shared fairly among the other sessions.

Proportionally fair allocations approximate max-min fair ones. In a proportionally fair allocation, any change in the rates results in the sum of the proportional changes being negative. Proportional fairness is better understood as the solution to a convex optimization problem whose objective function is {\em aggregate utility} (that is, the sum of the utilities of transport sessions) subject to capacity constraints provided that the utility functions are logarithmic. TCP congestion control is a distributed algorithm that induces a rate allocation, which is approximately proportionally fair. In this perspective, TCP is an algorithm for optimizing {\em social welfare} in that it solves a global optimization problem with aggregate utility in the objective function. It is, therefore, natural to ask: What makes such an algorithm stable in a game-theoretic sense?

\subsection{Rate allocation game}
\label{woeiurtyeirouty}

Toward the end of answering the previous question, we formulate a game-theoretic version of the rate allocation problem. Our formulation is intended to capture a {\em particular} scenario of the general rate allocation problem, namely, one where multiple similar transport sessions (similar in that they are transferring the same or similar content) of unknown duration simultaneously compete for the bandwidth of a bottleneck link, and our goal is to explain why these transport sessions would use TCP. Since we are focusing on a particular scenario, we do not aim to propose a general theory of TCP's game-theoretic stability (which would need to simultaneously explain more than one such scenarios), but rather to prove that even in this simple setting, players must be sophisticated enough to use TCP, which rules out the possibility of them being myopic.

To define any game, we need to specify the set of players, their strategies, and the payoffs that combinations of strategies yield.

\subsubsection{Players}

The vast majority of Internet traffic is content retrieval, and the decision on the protocol by which to retrieve content is made jointly by the content distribution system (e.g., BitTorrent) or content provider (e.g., Google or various ISP's) and the enduser. We, thus, assume that the players in the rate allocation game are {\em transport sessions} since protocol decisions are made at this level of granularity; for example, TCP congestion control ensures approximate proportional fairness at the granularity of TCP sessions. Let $N$ be the number of players, and we use $i$ as the running index of a player.

\subsubsection{Strategies}

The rate allocation game is a supergame consisting of an infinite sequence of identical stage games. At each stage game, we assume that each player has a binary choice of either using TCP (action $C$) or a custom transport protocol that obviates congestion control (action $D$). A strategy of player $i$ in this supergame is a sequence $\{s_i^t\}$, where the $s_i^t$ are functions mapping the period-$t$ history of play onto the action space $\{ C, D \}$ of the stage game. We denote the strategy of always cooperating by $C^{\infty}$, and the strategy of always defecting by $D^{\infty}$.

\subsubsection{Payoffs}

We assume that players compete for the bandwidth of a {\em single link;} if players are able to sustain cooperation in this environment, sustaining cooperation in environments where player competition is less fierce would pose no additional difficulties. 

Otherwise the payoff structure is that of a {\em symmetric prisoner's dilemma supergame}~\cite{Taylor}. The supergame is symmetric in that at each stage game a player's payoff depends only on the player's own choice, and the number of other players choosing $C$. That payoffs are symmetric in the event that all players cooperate is due to proportional fairness and the pedantic assumption that, in the absence of congestion control, each session would send at a rate higher than one $N$th of the link's capacity. That payoffs are symmetric among players that have defected is due to all such players using the same transport protocol and there is no reason to believe that congestion collapse would favor one defecting player over another.

Let $f_k$ be a player's payoff if the player chooses $C$ and $k$ others choose $C$ and let $g_k$ be the corresponding payoff if the player chooses $D$ and $k$ others choose $C$. It is assumed that:
\begin{enumerate}

\item $\forall k \geq 0, g_k > f_k$

\item $f_{N-1} > g_0$

\item $\forall k > 0, g_k > g_0$

\end{enumerate}
The first assumption captures that the choice to defect (i.e., to obviate congestion control) dominates cooperation. Indeed the first player to defect seizes a disproportionate fraction of the bandwidth, which he has to share with other defecting players at a much lower efficiency level than if everyone cooperated; that the efficiency level of a system that obviates congestion control is lower is attributed to the ensuing congestion collapse and is captured in the second assumption. The third assumption captures that the payoff of players using the aggressive transport protocol decreases as more players defect.

\subsubsection{Numerical example}

\begin{figure}[tb]
\centering
\includegraphics[width=7.5cm]{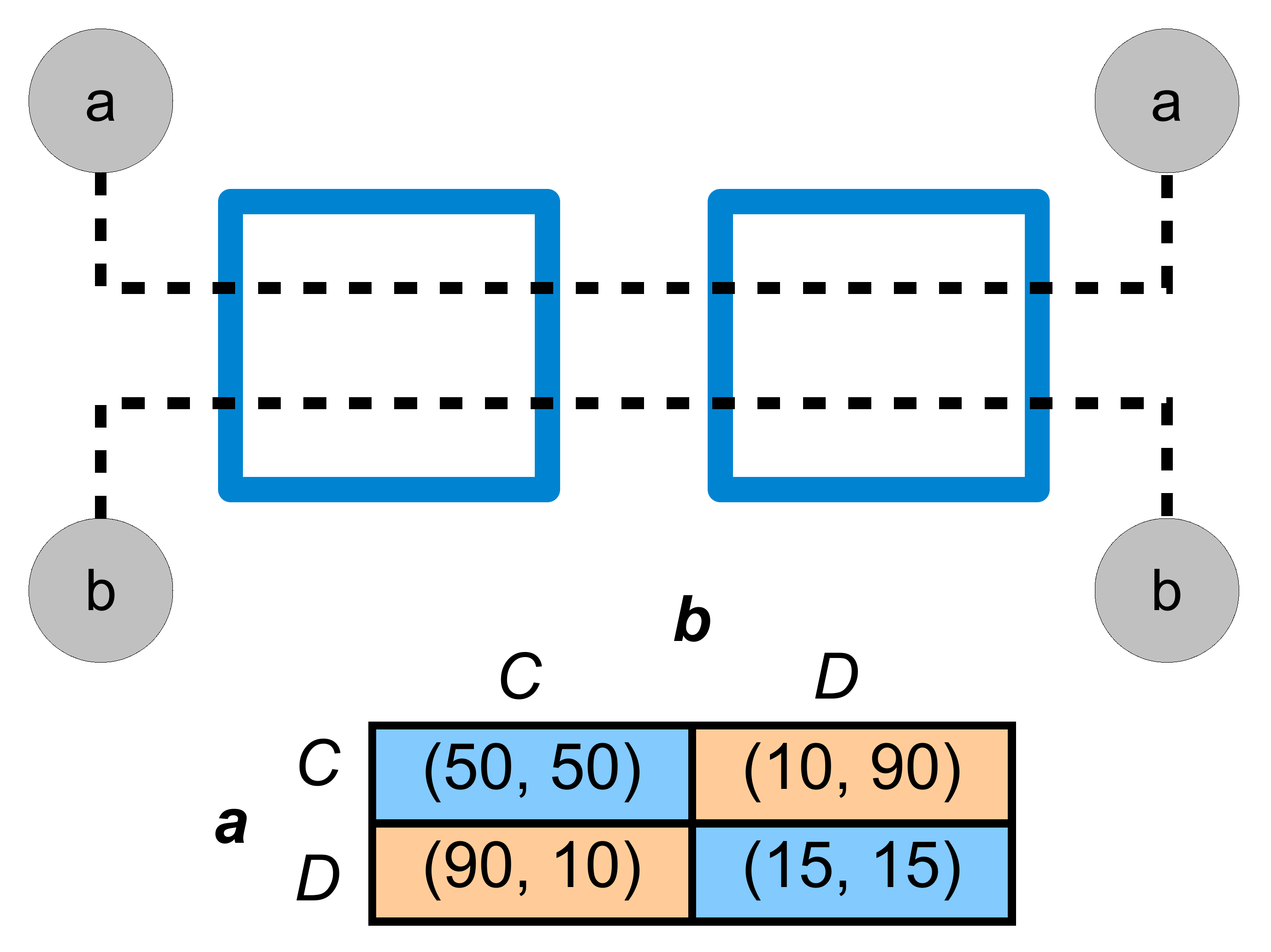}
\caption{\label{saldkjfnalxdkhjf}
Two flows crossing a link and one stage of the corresponding supergame.}
\end{figure}

Suppose the link has bandwidth of 100 units and serves flows $a$ and $b$ (as shown in Fig.~\ref{saldkjfnalxdkhjf}), which correspond to the players. Either player may choose to cooperate ($C$) or defect ($D$). If both cooperate, then the bandwidth is split evenly among them, and each gets a share of 50 units. If either player uses $C$ and the other uses $D$, the former receives 90 units of bandwidth whereas the latter only 10, and if both use $D$, because of congestion collapse, each receives 15 units. This interaction can be represented as the two-person prisoner's dilemma game (Fig.~\ref{saldkjfnalxdkhjf}). 

Prisoner's dilemma has the property that its unique equilibrium is for both players to choose $D$, that is, to obviate congestion control, however, this property no longer holds in the corresponding supergame, which has more equilibria.

\if(0)

\begin{figure}[]\hspace*{\fill}
\begin{game}{2}{2}
      & $C$    & $D$\\
$C$   &$50,50$   &$10,90$\\
$D$   &$90,10$   &$15,15$
\end{game}\hspace*{\fill}
\caption[]{\label{lskdjfhsdkjfh} Rate allocation game example.}
\end{figure} 

\fi

\subsection{TCP as an equilibrium}

The empirical outcome of play in the Internet can be approximated as {\em universal cooperation:} all players cooperate in all stage games. However, it is easy to show that universal {\em unconditional} cooperation corresponding to the strategy profile $$(C^{\infty}, \ldots, C^{\infty})$$ is {\em not} an equilibrium. To explain the empirical outcome we must, therefore, consider {\em conditionally cooperative strategies}, that is, strategies where cooperation is contingent upon previous outcomes. 

The most prominent of these strategies is {\em tit-for-tat}, which was originally proposed in the setting of two-person prisoner's dilemma supergames~\cite{Axelrod}. Tit-for-tat is to cooperate in the first game and to choose the opponent's previous strategy in succeeding games. The natural generalization of tit-for-tat in corresponding $N$-person games is to cooperate in the first game, and then cooperate if and only if at least $n$ players cooperated in the preceding game~\cite{Taylor}; call this strategy $B_n$. If every player uses $B_n$, then the supergame's outcome is indeed universal cooperation, however, is $(B_n, \ldots, B_n)$ an equilibrium? Taylor~\cite{Taylor} answers this question in the affirmative provided that $n = N-1$ and
\begin{align*}
\alpha_i \geq \frac{g_{N-1} - f_{N-1}}{g_{N-1} - g_0},
\end{align*}
where $\alpha_i$ is the discount factor of player $i$. That the discount factor must be large enough implies that to sustain cooperation it is necessary that players are {\em not myopic}, which corresponds with intuition: Myopic players focused on short-term payoffs cannot escape the dilemma of the one-shot game.

We have, therefore, identified a game and an equilibrium in this game such that if players use the corresponding equilibrium strategies, the outcome is consistent with the empirical outcome of transport sessions using TCP universally. This analysis is the first attempt to model rate allocation as a problem of technological competition between TCP and a more aggressive variant, and, although we have focused on one particular (but realistic) scenario, we are able to conclude that a general theory of TCP's architectural stability cannot but be based on conditionally cooperative strategies.

What's more, tit-for-tat is, to the extent of our knowledge, the {\em simplest} of all strategies that can sustain cooperation as an equilibrium, which implies that the Occam's razor singles out this strategy as the most likely explanation of the behavioral phenomenon we are trying to understand. We, may, thus, postulate that Internet players, and, in particular, content providers, do not defect from TCP congestion control {\em thinking} that defection will be counteracted with retaliation from other content providers. Papadimitrou was, therefore, in part, correct in predicting that ``If we see Internet congestion control as a game, we can be sure that its equilibrium is not achieved by rational contemplation, but by interaction and adaptation \ldots'' as tit-for-tat is indeed an adaptive strategy, however, we saw that it is rational contemplation (and a {\em theory of mind}) averting players from defecting.

Tit-for-tat is not the only equilibrium of the rate allocation game; for example, unconditional defection corresponding to the strategy profile $$(D^{\infty}, \ldots, D^{\infty})$$ also is. It is then natural to ask why the Internet population selects the cooperative equilibrium. We believe that TCP is a {\em focal equilibrium}~\cite{Myerson, Schelling}, that is, an equilibrium that is conspicuously distinctive from other equilibria and, therefore, one that everyone expects to be the outcome, and, hence fulfill, and there are many reasons for this: (1) TCP has been in stable operation for over 25 years, and it is known to work. (2) TCP is efficient, maximizing aggregate social utility, and equitable in the sense that it provides a level playing field; content providers would have abandoned it had it not facilitated fair competition. (3) The equilibrium of universal defection leads to congestion collapse, making it conspicuously unattractive; the Internet population would prefer to steer away from such an equilibrium.

\subsection{TCP and incremental deployability}

\subsubsection{Internet players are not myopic}

Since the Internet is to a large extent an anarchy, understanding its evolutionary forces is a necessary step toward innovation being effected. The previous analysis furthers this understanding by substantiating that, in the Internet, players are {\em not myopic} and that instead they have a {\em theory of mind}, which is an ability to think about other players' intentions: Although each provider has the myopic choice of employing an aggressive transport protocol to realize short-term benefits, they apprehend that acting on that choice would lead the way to all providers losing their business as other providers will follow suit to avoid conferring an advantage to a competitor. It is this theory of mind (and {\em not} altruism) that gives rise to cooperation and averts the long run losses that would otherwise ensue for all players. 

\subsubsection{Incremental deployability is not panacea}

The networking community has been swift in recognizing the need to understand the Internet's evolutionary forces, and the push toward incrementally deployable technologies seems to spring out of this understanding. Incremental deployability is understood a property a technology may or may not have, and if it is the former, it implies benefits to the early adopters of the technology. From this definition it is not immediately apparent that building incrementally deployable technologies will help innovation efforts as it is not apparent that such technologies will receive deployment traction. For example, that early adopters benefit does not necessarily imply that the early adopters (or anyone else for that matter) won't lose as adoption gains ground. Should early adopter benefits matter more than long run losses? Naturally, they shouldn't, and, in fact, they don't: In TCP, unilateral defection is incrementally deployable against cooperation, however, as discussed earlier, players choose not to defect to avoid the losses that transpire thereafter. This further implies that incremental deployability is not panacea in the efforts to effect innovation, and that the community should think about innovation more broadly.

\subsubsection{Incremental deployability's niche}

That is not to say, however, that incremental deployability is not an important concept; if, for example, a technology benefits the social good it is all the better if that technology is incrementally deployable. There are many examples of technologies that benefit the social good at the network and transport layers of the Internet that have been notoriously hard to deploy, and we believe that incremental deployability sprang out of folk wisdom to address the need of deploying these technologies. Since these technologies are also {\em inherently} difficult to deploy incrementally, at the initial stages of this research we were tempted to attribute the deployment failures to the lack of incremental deployability. However, TCP's dominance at the transport layer rules out this possibility: Internet players are not myopic, and to understand why innovation is failing we need to look elsewhere. In the rest of this paper, we attempt a first-principles approach to understanding and overcoming these deployment failures.

\section{Why is innovation failing?}
\label{iwueyrajshdgflgkhj}

Superior network- and transport-layer technologies whose success depends on positive externalities are notoriously difficult to deploy. In this section, in trying to understand {\em why}, we devise a game-theoretic model known as a {\em stag hunt} that captures the salient characteristics of the deployment failures. Using game-theoretic reasoning we reach the conclusion that the lack of deployment is attributed to players being {\em risk averse}---they are reluctant to invest fearing that other players will not follow suit---a conclusion that corresponds with simple intuitive reasoning. This corroborates the claim that the stag hunt succeeds in capturing the salient characteristics of the phenomenon we're scrutinizing. Having reframed the problem in game-theoretic terms opens new possibilities for attacking it, a subject we take on in Section~\ref{qwioerqwerjhejkl}.

\subsection{Case studies}

The network layer of the Internet is that part of the architecture which provides a best-effort end-to-end packet delivery service out of a collection of routers and links. The main protocols operating at this layer are IP, intradomain routing protocols such as OSPF and IS-IS, the interdomain routing protocol, namely, BGP, and (to some extent) DNS. There have been many attempts to effect innovation at the network layer the most prominent of which are the efforts to deploy IPv6, S-BGP, IP multicast, IP QoS, and DNSSEC. Similarly there have been many attempts to effect adoption of superior to TCP Reno (the dominant transport-layer algorithm) transport-layer technologies such as TCP Vegas and FAST TCP. We claim that it is possible to capture the salient characteristics of these deployment efforts (which we individually discuss below) in one simple mathematical model. However, since DNS and DNSSEC are mostly application-layer protocols we defer their discussion in the interest of space.

\subsubsection{IPv6}

Motivated by the exhaustion of the $32$-bit IPv4 addresses, IPv6~\cite{Loshin} provides a $128$-bit address space (and other secondary features). In all detail, IPv6 is an incremental change to IPv4, however, since IP is ubiquitous in the Internet, adoption of IPv6 requires upgrades at hosts and routers. 

Rather than upgrading to the new protocol, the Internet population has manifested strong preference for temporary point solutions such as NAT (Network Address Translation)~\cite{NAT} and address trading \cite{Beijnum}, whose success does not hinge on universal adoption, which is, in fact, the only apparent reason that the Internet population shows a preference against IPv6 since NAT is arguably technically inferior, violates basic architectural principles, and induces significant complexity. (See~\cite{WD} for a discussion.) However, the reasons that conditioning success on universal adoption is disconcerting enough to drive the Internet population to technically inferior choices are narrowly understood.

To the extent of our knowledge, there is no comprehensive plan to effect adoption of IPv6, and most efforts and proposals focus on increasing the benefits of partial adoption using techniques such as IPv6-over-IPv4 tunnels~\cite{Loshin} and anycast~\cite{RSM}. Although the ultimate goal of these techniques is clearly to eliminate deployment barriers (i.e., to convert IPv6 into an incrementally deployable technology), this goal has remained elusive.

\subsubsection{S-BGP}

Routing attacks such as {\em prefix hijacking} can compromise the availability, integrity, and privacy of Internet communication. S-BGP~\cite{Kent} aims to protect the Internet from routing attacks and to that end it requires making use of a public key infrastructure (PKI) as well as changes to the BGP protocol to cryptographically protect BGP messages. Adoption of S-BGP entails building a PKI and upgrading BGP-speaking routers in autonomous systems. The Internet population has been reluctant to adopt S-BGP, and the US government has considered the possibility of mandating its deployment.

Gill et al.~\cite{Gill} propose and evaluate a sweeping plan for the adoption of S-BGP (or soBGP~\cite{soBGP}), which does not succeed, however, in converting S-BGP to an incrementally deployable protocol: The success of their scheme depends on {\em exogenous pressure} (such as regulation or subsidies) for adoption to gain initial deployment momentum.

\subsubsection{IP Multicast}

IP multicast makes {\em group communication} (whether from one source to many destinations or from many sources to many destinations) efficient and to that end it requires extensions to intradomain and interdomain routing protocols~\cite{Deering, Rosenberg}. Intradomain multicast has received limited adoption by some service providers, however, despite having been the subject of extensive research since the early '90s, interdomain multicast has not received any adoption.

\subsubsection{IP QoS}

QoS provides quantitative performance guarantees for end-to-end traffic and allows users to select the level of service. The most prominent efforts to extend IP's best-effort service model and provide QoS in IP networks are IntServ and DiffServ, the former being a fine-grained mechanism operating at the granularity of flows and the latter being a coarse-grained mechanism operating at the granularity of traffic classes. Both IntServ and Diffserv require from every router to participate in the system.

\subsubsection{Transport-layer innovation}

Since the '90s, researchers have been trying to improve on TCP congestion control's performance. Much effort has focused on {\em delay-based congestion avoidance algorithms}. The first algorithm of this kind was TCP Vegas~\cite{TCP-Vegas}.  TCP Vegas and related algorithms such as FAST TCP~\cite{FAST-TCP} (see~\cite{RFC6297} for a survey) are based on a different principle than TCP Reno: Instead of detecting incipient congestion using packet losses, they use round-trip-time measurements to that end (hence the name delay-based algorithms). Delay-based algorithms improve network throughput if universally adopted, however, if delay-based and loss-based algorithms compete, the latter capture a disproportionate fraction of the bandwidth as they are more aggressive.

\subsection{Modeling innovation as a stag hunt}

The success of the previous technologies depends on positive externalities in that benefits increase with the number of adopters, and none is known to be incrementally deployable. In fact, technologies whose success depends on positive externalities are inherently difficult to deploy incrementally.

To model competition between an incumbent and a superior emerging technology whose success depends on positive externalities we use the game-theoretic model of the {\em stag hunt}, which Skyrms~\cite{Skyrms} lucidly defines as follows: ``Let us suppose that the hunters [in a group] each have just the choice of hunting hare or hunting deer. The chances of getting a hare are independent of what others do. There is no chance of bagging a deer by oneself, but the chances of a successful deer hunt go up sharply with the number of hunters. A deer is much more valuable than a hare. Then we have the kind of interaction that is now generally known as the stag hunt.''

In our setting, deer corresponds to the emerging technology, hare corresponds to the incumbent, and the hunters correspond to the potential adopters. That the chances of getting a hare are independent of what others do reflects the often realistic assumption that the incumbent technology neither benefits nor suffers from adoption of the emerging technology. That there is no chance of bagging a deer by oneself reflects that the emerging technology is not incrementally deployable, that the chances of a successful deer hunt go up sharply with the number of hunters reflects the existence of positive externalities, and that a deer is much more valuable than a hare reflects that the emerging technology is superior to the incumbent. What follows is a more in-depth discussion of our assumptions and of the stag hunt.

\subsubsection{Modeling network-layer innovation}

We think of network-layer players as corresponding to the administrative authorities of {\em autonomous systems}, and we use $i$ as the running index of a player. We assume that each player has a choice of either adopting the new technology (strategy $A$) or defecting, that is, seceding from adoption (strategy $D$). In contrast to the previous section, viewing network-layer innovation as a supergame does not provide further insight into the problem; the empirical outcome of the game is universal defection, which can be explained in the one-shot game without resorting to modeling repeated interactions.

We assume that the payoff of a player who adopts depends on the size and composition of the connected component of adopters in the autonomous system graph to which the player's autonomous system belongs, and that this payoff goes up sharply with the size of this connected component. The payoff of a player who defects is zero. If the size of an adopter's connected component is small enough, the adopter's payoff may be negative (meaning that the investment effort costs more than the benefit).

If we assume that {\em all} players must upgrade to the emerging technology to make it effective, the payoff function is given by the following simple formula:
\[
  u_i(s_i, s_{-i}) = \left\{ 
  \begin{array}{l l}
    \beta_i-\gamma_i, & \text{if $s_i=A$ and $q(s_{-i}) = 1$}\\
    -\gamma_i, & \text{if $s_i=A$ and $q(s_{-i}) = 0$}\\
    0, & \text{if $s_i=D$.}\\
  \end{array} \right.
\]
where $\beta_i-\gamma_i > 0$ is the net benefit of adoption, $\gamma_i > 0$ is the investment cost, and $q(s_{-i}) = 1$ if and only if there is no $D$ in the vector $s_{-i}$.

In fact, if we were to make this additional simplifying assumption, we would not significantly lose generality as players need not correspond one-to-one with the potential adopters. Rather these players can represent a good enough {\em subset} of the potential adopters such that if the players in the subset adopt, the emerging technology becomes incrementally deployable. However, this would only be an approximation of the incentive structure of the game, and the reader should bear both formulations in mind.

\subsubsection{Modeling transport-layer innovation}

Competition between delay-based and loss-based transport-layer algorithms can be captured by the rate allocation game of Section~\ref{woeiurtyeirouty} where cooperation corresponds to using a delay-based algorithm and defection corresponds to using the loss-based one. The rate allocation game is a supergame, however, it is possible to simplify matters if we limit the supergame's strategies to tit-for-tat and unconditional defection, in which case, the rate allocation game becomes a stag hunt (for example, see~\cite{Taylor, Skyrms}).

\subsubsection{Properties of the stag hunt}

The stag hunt has two pure-strategy Nash equilibria, namely, universal adoption and universal defection. Since the former yields a positive payoff to each player whereas the latter a payoff of $0$, universal adoption is a Pareto dominant equilibrium and rational choice would predict selection of this equilibrium. However, to remain consistent with empirical evidence, we have to accept the Pareto inferior equilibrium of no adoption as the outcome of the game. Selection of the Pareto inferior equilibrium is generally referred to in the literature as a {\em coordination failure}~\cite{Cooper}, which we seek to explain and overcome.

\subsection{Explaining the coordination failure}

Observe that the coordination failure {\em contradicts rationality} in that potential adopters are given the opportunity to benefit both individually and as a group, however, they reject it. Observe further that the Pareto dominant equilibrium is not incrementally deployable, which would explain the deployment failure granted that players are myopically selfish, a hypothesis we are in the position to rule out: In the rate allocation game of Section~\ref{woeiurtyeirouty}, players were shown to have a theory of mind and there is no reason to believe that in the innovation games we are considering players are devoid of such capability. To explain the coordination failure we attempt to dig deeper into the psychology of human behavior. To that end we look first at experimental evidence.

\subsubsection{Experimental evidence}

``But there is contemporary experimental evidence that people will sometimes hunt stag even when it is a risk to do so. \ldots \hspace{0.1 mm} If the group starts in the basin on attraction of stag hunting, then the group almost always converges to all stag hunters. If the initial composition of the group is in the basin of attraction of hare hunting, hare hunters take over''~\cite{Skyrms}. Unfortunately, insofar innovation whose success depends on positive externalities is concerned, potential adopters qualify as {\em hare hunters}. To the extent of our knowledge, the only innovation facing a stag hunt whose deployment succeeded using the Internet's evolutionary forces alone was TCP congestion control.\footnote{It is worth noting that the deployment of TCP/IP had been mandated~\cite{Abbate} whereas BGP replaced a significantly more limited predecessor.} (TCP's success is discussed further at the end of this section.) Since the remaining failures contradict rationality, our explanation by necessity resorts to theories of {\em bounded rationality}~\cite{Gilboa, Wilkinson}.

\subsubsection{Risk aversion}

Which equilibrium will be selected depends on what each player believes others will do. Why would a player doubt though that others will adopt? If all players are rational and this is common knowledge, there is no reason to doubt that others will adopt as adoption is the Pareto superior outcome. However, if rationality is {\em not} common knowledge in that some players doubt that other players are rational, then adoption becomes a {\em risky} undertaking vis-\`a-vis defection (as defection guarantees the status-quo payoff irrespective of the other players' choices). 

Investment decisions at the level of autonomous systems, for example, in the Internet are made by self-interested organizations, which are unlikely to be grossly irrational (to the extent of acting against their own interests), and this is generally known. However, the {\em history} of repeated failures to effect innovation reinforces doubt  that other organizations will follow suit in an investment/deployment effort.

Still, the previous arguments do not suffice to explain the coordination failure as the benefit of adoption may outweigh the risk. However, it has been shown experimentally that the presence of risk in a choice may introduce {\em significant} bias in human subjects against making that choice, a phenomenon that we refer to as {\em risk aversion}. 

Research in the psychology of decision making under risk by Kahneman and Tversky~\cite{Kahneman} shows that individuals have a general preference for outcomes that are obtained with certainty over outcomes that are merely probable even if in expectation the probable outcomes yield higher benefits. Since players that are choosing between the incumbent and the emerging technology are, in fact, choosing between an outcome that is obtained with certainty and an outcome that is merely probable, risk aversion {\em biases} players against the Pareto superior outcome.

There has also been an attempt to mathematically capture how the presence of risk affects decision making in {\em coordination games} (of which the stag hunt is a special case). To that end Harsanyi and Selten~\cite{HS} propose the solution concept of {\em risk dominance:} Of all equilibria, the one being selected is that which has the smallest {\em risk factor} (that is, is less risky), a quantity that depends on the payoff structure of the game. In the stag hunt, unless the benefit of adoption is very large, the inferior equilibrium risk dominates the superior equilibrium, and their solution concept predicts selection of the former in agreement with empirical evidence.

We may conclude that superior technologies whose success depends on positive externalities are notoriously difficult to deploy {\em because it is risky to do so,} and humans have a natural propensity against choosing risky prospects. Our conclusion points to {\em risk management} as being essential for deployment success. This conclusion corresponds with intuition: Incrementally deployable technologies (for example, NAT) are effective in risk management as their success does not entail any (short-term) risk. Motivated by this conclusion, we look into the possibility of a standard risk management strategy, namely, {\em insurance}, as a deployment method, however, before taking on this subject in Section~\ref{qwioerqwerjhejkl}, we revisit the transport-layer game of Section~\ref{woeiurtyeirouty}.

\subsection{Discussion}
\label{owiuredmvcnx}

In both the transport- and network-layer games, players face a choice between cooperation and defection, the payoff structure is quite similar, and the players themselves are also quite similar (they reason about other players' intentions and they are risk averse). We are, therefore, interested in answering why there is a discrepancy in outcome in just one particular game, the one where TCP competes with transport sessions that obviate congestion control. We believe there are three reasons for this: 
\begin{itemize}

\item {\em Players evaluate other players' willingness to cooperate in divergent ways because of the history of play:} Although loss-based algorithms have been in stable operation for over 25 years, delay-based algorithms and network-layer technologies have suffered repeated deployment failures. Therefore, players in the former game start in the basin of attraction of stag hunting and stag hunters take over whereas players in the latter games start in the basin of attraction of hare hunting and hare hunters take over. 

\item {\em Contemporary innovation faces higher deployment barriers because of the size of the Internet:} Congestion control was deployed in a much smaller Internet. 

\item {\em Lack of cooperation against obviating congestion control leads to congestion collapse:} Players that face a problem of choice between two prospects, one risky and the other certain, use both the probability of the risky prospect materializing and the payoff difference in their decision. In particular, since the lack of cooperation at the transport layer leads to congestion collapse, the expected payoff of cooperation is significantly higher than the payoff of defection, and, therefore, players choose to cooperate despite cooperation being a risky undertaking.

\end{itemize}

\section{Engineering evolution}
\label{qwioerqwerjhejkl}

We have hitherto tried to understand cooperation as an evolutionary force in the Internet, and have found that evolution's outcome may be undesirable. In this section, we ask whether we can intervene in evolution's course toward the end of effecting desirable outcomes.

\subsection{Problem formulation}

The problem we aim to solve in this section is how to induce deployment of a superior emerging technology whose success depends on positive externalities. Using our game-theoretic formulation of this problem, we are equivalently seeking to advance a group of risk-dominated players playing a stag hunt to the Pareto dominant equilibrium.

\subsection{Using insurance against risk aversion}

The idea we explore to conciliate risk aversion is to offer the possibility of {\em insurance}: Players can buy an insurance policy from an insurance carrier to hedge against the risk of a futile investment in the emergent technology. There are many conceivable ways to formulate the insurance policy that should naturally depend on which technology is being promoted for adoption, however, devising insurance policies for specific technologies is beyond the scope of this paper. Notwithstanding the differences between technologies, we have demonstrated significant commonalities, which we leverage to demonstrate that insurance can be an effective deployment method.

Toward the end of demonstrating the effectiveness of insurance, we consider a simple insurance policy that covers the benefit of {\em universal} adoption, under the condition that the insured player adopts. In the following, we {\em prove} that such a simple insurance policy is effective in incentivizing the players to adopt the emerging technology even if universal adoption is {\em necessary} for adopters to reap the benefits of the emerging technology. The insurance policy further specifies a {\em target date} in the future when eligible players are compensated unless adoption has already taken place. This insurance policy changes the incentive structure of the environment as follows.

\subsubsection{Players}

The player set is as before---it may correspond to administrative authorities of autonomous systems or to transport sessions.

\subsubsection{Strategies}

Each player has a choice of three strategies (instead of two), which are to adopt ($A$), to adopt with insurance ($B$), and to secede/defect ($D$).

\subsubsection{Payoffs}

The payoff function is defined by the following formula where $\beta_i$ is the benefit of the new technology (to player $i$), $\gamma_i$ is the investment cost, $\epsilon_i$ is the insurance premium, and $\delta_i$ is the reimbursement from the carrier to the player under an adoption failure:
\[
  u_i(s_i, s_{-i}) = \left\{ 
  \begin{array}{l l}
    \beta_i-\gamma_i, &  \text{if $s_i=A$ and $q(s_{-i}) = 1$}\\
    -\gamma_i, &  \text{if $s_i=A$ and $q(s_{-i}) = 0$}\\
    \beta_i - \gamma_i - \epsilon_i, & \text{if $s_i=B$ and $q(s_{-i}) = 1$}\\
    \delta_i - \gamma_i - \epsilon_i, & \text{if $s_i=B$ and $q(s_{-i}) = 0$}\\
    0, &  \text{if $s_i=D$.}\\
  \end{array} \right.
\]
where $q(s_{-i}) = 1$ if and only if there is no $D$ in the vector $s_{-i}$. 

In words, a player $i$ that adopts receives the benefit $\beta_i$ of the new technology as long as adoption happens and adoption happens as long as none of the players chooses $D$. If a player adopts and at least one other player secedes, then the player that adopts receives a negative payoff. However, if a player adopts with insurance, he is guaranteed a payoff of $ \delta_i - \gamma_i - \epsilon_i $, which should be positive, irrespective of the choices of the other players. (We assume here that player $i$ pays $\epsilon_i$ to the insurer, and if adoption happens, the player receives $\beta_i$ by making use of the new technology, whereas if adoption does not happen, the player receives $\delta_i$ by the insurer.) Finally, a player that secedes receives a payoff of $0$.

\subsection{Efficacy of insurance}

\subsubsection{Proposition}

We claim that the previous game has a unique pure-strategy equilibrium, namely, the Pareto dominant equilibrium of the corresponding stag hunt.

\subsubsection{Proof}

First, we show that strategy profiles containing one or more $D$'s cannot be equilibria. Indeed, in any such profile, a player using $D$ benefits by purchasing insurance (their payoff increases from $0$ to at least $\delta_i - \gamma_i - \epsilon_i$). Then, we show that strategy profiles containing no $D$'s and one or more $B$'s cannot be equilibria. Indeed, in any such profile, a player using $B$ benefits by giving up insurance (their payoff increases by $\epsilon_i$). Finally, $(A, \ldots, A)$ is a Nash equilibrium as no player benefits by unilaterally deviating to another strategy, which completes the proof.

\subsubsection{Discussion}

Note that the status quo outcome $(D,\ldots,D)$ is no longer an equilibrium, and that each player's best response to the status quo is to adopt the emerging technology {\em with insurance} (strategy $B$). What's more,  $(B,\ldots,B)$ is not an equilibrium either as starting from this outcome each player benefits by adopting the emerging technology {\em without insurance} (strategy $A$). Together, $(A,\ldots,A)$ becomes incrementally deployable against the status quo $(D,\ldots,D)$ in {\em two} incremental steps. However, since the first of these steps must happen mentally in the minds of the potential adopters, it would not be surprising if in experimental settings some players would be unwilling to take this mental step and choose $B$ instead. Therefore, it would not be surprising for outcomes where some players choose $B$ and others choose $A$ to emerge in experimental settings. 

Note also that the reimbursement amount $\delta_i$ must be large enough such that $\delta_i - \gamma_i - \epsilon_i > 0$, and that it is not necessary to precisely estimate the benefit $\beta_i$ of the emerging technology. Note, finally, that the insurer does not lose by offering insurance, and may also benefit if some players choose $B$.

\section{Related work}

This paper is a {\em game-theoretic approach} to understanding the {\em Internet architecture} and to effecting {\em architectural innovation} in the Internet infrastructure at the network and transport layers.

\subsection{Internet architecture}

\subsubsection{Understanding protocol design}

Internet protocols such as TCP or BGP were designed in an ad hoc manner, and the effort to devise rigorous models of TCP (e.g.,~\cite{Kelly2, Low, Understanding-TCP-Vegas, Shorten}) or BGP~\cite{Griffin} has received prominent attention. However, these models explain {\em what} TCP, for example, does rather than {\em why} TCP Reno is chosen by the Internet population or {\em why} TCP Vegas is not adopted.

\subsubsection{Theorizing about the Internet architecture}

Chiang et al.~\cite{Decomposition} theorize that network architecture can be understood as an asynchronous distributed algorithm solving a global optimization problem with network utility in the objective function. However, this theory does not justify why the Internet population is eager to follow the algorithmic steps. For example, despite IPv6's benefits, the Internet population has been reluctant to adopt it.

\subsubsection{Explaining emergent phenomena}

Willinger and Doyle~\cite{WD} argue that the evolution of the Internet follows a spiral of increasing complexity to achieve robustness to sensitivities. We believe that much of this complexity is imputable to a lack of understanding of the Internet's evolutionary forces: Understanding these forces better will enable us to transition to principled, easier to understand architectures removing unnecessarily complex designs necessitated by point solutions.

Akshabi and Dovrolis~\cite{Hourglass} propose a rigorous model of protocol competition in a layered architecture, which they use to explain the Internet hourglass using simulation. The basis of their model, dubbed {\em EvoArch}, is a directed acyclic graph where nodes represent protocols and links represent {\em protocol dependencies}. In all detail, {\em EvoArch} is a technology-centric model of evolution where user incentives are only captured implicitly by the ``evolutionary value'' that dependencies confer on protocols. In contrast, users are principal entities in our model of evolution, and user incentives are captured in their generality by player utility functions.

Diot et al.~\cite{RIP-multicast} and Crowcroft et al.~\cite{RIP-QoS} provide explanations as of why service providers forwent deployment of the much anticipated multicast and QoS architectures respectively. The perspective that we offer in this paper that service providers are forgoing deployment because of risk is complementary to those explanations.

\subsubsection{Effecting innovation}

There are many proposals aiming to confront the Internet's technology-centric barriers to innovation (for example, see~\cite{Impasse, FII, Ghodsi, XIA}). In contrast, this paper confronts the system's human-centric barriers. Previous efforts to effect adoption of technologies whose success depends on positive externalities (e.g.,~\cite{RSM, Gill}) facilitate a smoother transition from the incumbent to the emerging technology, but do not eliminate deployment barriers.

\subsection{Diffusion of innovation}

Diffusion models of innovation assume that the adoption of an innovation by an agent in a social system (typically represented by a graph) is contingent {\em only} on adoption by enough agents in this system. Examples of diffusion models include ones where an agent adopts if enough neighbors have adopted~\cite{Spread} or if the agent belongs to a large enough connected component of adopters~\cite{Goldberg2}. In contrast, we assume that the adoption of an innovation is contingent on it being a {\em best response} to the choices of other agents. Therefore, agents in our model {\em optimize} whereas in diffusion models {\em satisfice}. Furthermore, our model is more general in that it allows modeling competition among an arbitrary number of technologies.

Researchers have used diffusion models to look for the smallest possible set of agents ({\em seedset}) such that if exogenous pressure causes them to adopt, the entire social system adopts (e.g.,~\cite{Spread, Goldberg2}). The advantage of this method is that it concentrates exogenous pressure, however, there is no guarantee it will find a small enough seedset. In contrast, insurance makes adoption individually beneficial for each agent.

\subsection{Game Theory}

\subsubsection{Algorithmic Game Theory}

The effort to understand the Internet from the perspective of Algorithmic Game Theory has focused on efficiency questions such as `What is  the price of anarchy?' (For example, see~\cite{PriceOfAnarchy, HowBadisSelfishRouting, selfish_routers}.) That the Internet is an anarchy raises, however, equally important, if not more, questions such as `What are the Internet's evolutionary forces?' and `How can we effect innovation?' Our effort in this paper has focused on answering those latter questions.

\subsubsection{Evolutionary Game Theory}

Our model of architectural evolution is closest to the Maynard Smith and Price's game-theoretic model of biological evolution~\cite{TheLogicOfAnimalConflict, Evolution}. Their model is similar to ours in that society's agents correspond to {\em organisms}, technologies correspond to {\em behaviors}, and the question is which strategies survive competition. However, in their model, behaviors are genetically inherited, which gives rise to the {\em evolutionary stable strategy} as the equilibrium concept. In contrast, since in our model agents are humans, the relevant equilibrium concepts can be much more sophisticated (and may even correspond to equilibrium concepts in supergames as in TCP).

\subsubsection{Coordination games}

The method of using insurance to advance players from the inferior to the superior equilibrium in a stag hunt is related to a method by Cooper~\cite{Cooper} to the same effect. In the two-player coordination games Cooper considers, he allows one player the option of receiving a sure outcome instead of playing the coordination game and then reasons that subject to a technical condition this leads to selection of the Pareto superior outcome, a conclusion that he also verifies experimentally. In contrast, we consider $N$-player games, insurance is an option available to all players, and it is not an ``outside option;'' once a player purchases insurance he commits to adoption.

\section*{Acknowledgments}

I'd like to gratefully acknowledge numerous discussions with Jon Crowcroft that significantly influenced the results in this paper. I am also thankful to Miltiades Anagnostou for an insightful discussion and to Jennifer Rexford for comments on earlier drafts.

\small

\bibliography{hotnets}

\end{document}